\documentclass[showpacs,preprintnumbers,amsmath,amssymb]{revtex4}

\usepackage{epsfig}
\usepackage{graphicx}
\usepackage{dcolumn}
\usepackage{bm}

\begin{document}

\preprint{BARI-TH 477/03}

\title{
Phase shifts of synchronized oscillators and the
systolic/diastolic blood pressure relation}
\author{L.Angelini$^{1,2,3}$, G. Lattanzi$^4$, R. Maestri$^5$, D. Marinazzo$^{1,2}$,
G.Nardulli$^{1,2,3}$, L. Nitti$^{1,3,6}$, M. Pellicoro$^{1,2,3}$,
G. D. Pinna$^5$,  S. Stramaglia$^{1,2,3}$}
 \affiliation{$^1$TIRES-Center
of
Innovative Technologies for Signal Detection and Processing,\\
Universit\`a di Bari, Italy\\
$^2$ Dipartimento Interateneo di Fisica, Bari, Italy \\
$^3$Istituto Nazionale di Fisica Nucleare, Sezione di Bari, Italy \\
$^4$Hahn-Meitner Institut, Abt. Theoretische Physik SF5, Glienickerstrasse 100, 14109
Berlin, Germany\\
$^5$Divisione di Cardiologia e Bioingegneria, Fondazione Salvatore
Maugeri, IRCCS
Istituto Scientifico di Montescano (PV), Italy \\
$^6$D.E.T.O., University of Bari, Italy}

\date{\today}

\begin{abstract}We study the phase-synchronization properties
of  systolic and diastolic arterial pressure in healthy subjects.
We find that delays in the oscillatory components of the time
series depend on the frequency bands that are considered, in
particular we find a change of sign in the phase shift going from
the Very Low Frequency band to the High Frequency band. This
behavior should reflect a collective behavior of a system of
nonlinear interacting elementary oscillators. We prove that some
models describing such systems, e.g.  the Winfree and the Kuramoto
models offer a clue to this phenomenon. For these theoretical
models there is a linear relationship between phase shifts and the
difference of natural frequencies of  oscillators and a change of
sign in the phase shift naturally emerges.

\end{abstract}

\maketitle
\section{Introduction\label{intro}}

Time series of physiological origin very often display synchronous behavior. Most likely,
this is the result of collective behavior of a huge number of nonlinearly interacting
elementary oscillators. Different examples of this phenomenon, as well as models of it,
can be found for example in \cite{winfree80}. In the present paper we address a further
example, i.e. the relation between diastolic (DAP) and systolic (SAP) blood pressure
signals in healthy subjects. The analysis of blood pressure fluctuations may provide
significant information on the physiology and pathophysiology of the autonomic control of
the cardiovascular function \cite{med1}, \cite{med2}. The synchronization of these
signals is expected, though  a detailed study of its features is apparently still
lacking. In a previous paper \cite{pinna} it was noticed that DAP and SAP are
characterized by a phase lag in the very low frequency band (VLF). The analysis of
\cite{pinna} uses Fourier analysis, which is not particularly useful when non-stationary
effects play a relevant role. In the present work we address two questions about the
DAP/SAP relationship: Is the phase lag depending on the frequency band? Is the phase lag
connected to a causal relation between SAP and DAP? To address these questions, we
measured DAP and SAP signal in a number of healthy subject. Studying the mutual
interdependency between the two time series, we conclude that there is not a causal
relationship between DAP and SAP time series, i.e. none of
     the two is driver for the other.
 Moreover, a significant phase delay is found,
  for healthy subjects, in the VLF
  band and in the high frequency (HF) band. The phase shift
   between DAP and SAP is positive in VLF band and negative
           in the HF band. This change of sign in the phase shift
           has its origin in the regulatory mechanisms of blood
           circulation. A physiological interpretation of these
           mechanisms is beyond the scope of our work; however the
           hypothesis that synchronization results from the
           collective behavior of elementary nonlinear oscillators
           may offer a clue to its understanding. To exploit this
           idea we use below two well known models of coupled
           oscillators, the Winfree model \cite{winfree67} and the Kuramoto model
           \cite{kuramoto}.

 Winfree's paper \cite{winfree67} on coupled oscillators
  provided one of the first tractable examples of a self-organizing system.
   Winfree introduced an approximation that has become
   the standard approach to the study of ensembles of biological
   oscillators: In the  weak coupling limit, amplitude variations
   could be neglected and the oscillators could be described only
    by their phases along their limit cycles. Winfree also discovered
    that systems of oscillators with randomly distributed frequencies
    exhibit a remarkable cooperative phenomenon,
    reminiscent of a thermodynamic phase transition, as the variance
    of the frequencies is reduced. The oscillators remain incoherent,
    each running near its natural frequency, until a certain threshold
    is crossed. Then the oscillators begin to synchronize spontaneously.
    Winfree model was subsequently
    modified by Kuramoto who provided an analytically solvable
    version of it \cite{kuramoto}. This field of study has been
very active all along and the analysis of synchronization
phenomena remains a
     thriving area of research, see  \cite{stroreview} for a
     review. Having in mind our experimental findings on the  SAP/DAP relation,
      we first examine in Section \ref{win}  the phase shift between
      coupled oscillators in these models, once synchronization has been reached. We
      observe that there exists a simple linear
dependence between phase shifts of synchronized oscillators and
the difference between their natural frequencies. This phenomenon,
to our knowledge never noticed before, offers a simple
            mechanism to describe the change of sign in the phase
             lag as the frequency band is changed. In Section
             \ref{phy} we describe the experimental data and
             analyze them using the theoretical approach of
             Section \ref{win}. Moreover we discuss the problem of
             the causal relation between the DAP/SAP time series.
             Finally in Section \ref{conclu} we draw our
             conclusions.
 \section{Phase shifts of synchronized oscillators\label{win}}
 \subsection{Winfree model \label{winfreemodel}}
  The Winfree model is defined by the set
of equations ($i=1,...N$)
\begin{equation}
 \dot\theta_i(t)=\omega_i+\frac{1}N\sum_{j=1}^N\,\kappa\,
P(\theta_j) R(\theta_i), \label{eq1}
\end{equation}It describes a set of $N\gg 1$ coupled non linear
oscillators, with coupling constant proportional to $\kappa$.
$\theta_i(t)$ is the phase of the $i-$th oscillator;
$\{\omega_i\}$ describes a set of natural frequencies taken
randomly from a distribution $g(\omega)$. $P(\theta_j)$ is the
influence function of the $j-$th; $ R(\theta_i)$ is the
sensitivity function giving the response of the $i-$th oscillator.
We shall assume below : $g(\omega)=1/2\gamma$ for $\gamma\in
[\omega_0-\gamma,\omega_0+\gamma]$, $g(\omega)=0$ otherwise. In
the previous equation $ P(\theta)=1+\cos\theta$,$\quad
R(\theta)=-\sin\theta$. The phase diagram of the Winfree model has
been recently discussed \cite{strogatz00}. In  particular the
long-time behavior of the system is characterized by a synchronous
dynamics for $\kappa$ and $\gamma$ not very large. For
$\omega_0=1$ synchronization occurs for $\kappa<0.77$ and
$\gamma<h(\kappa)$, where the function $h(\kappa)$ can be found in
Fig. 3 of Ref. \cite{strogatz00}; in any case $\gamma<0.2$. This
means that all the oscillators are characterized by a common
average frequency (or rotation number)
$\rho_i=\lim_{t\to\infty}\theta_i(t)/t$. The Winfree model can
describe different sets  of pulse-coupled biological oscillators,
see e.g. \cite{walker,buck,peskin}.

\begin{figure}[ht!]
\begin{center}
\epsfig{file=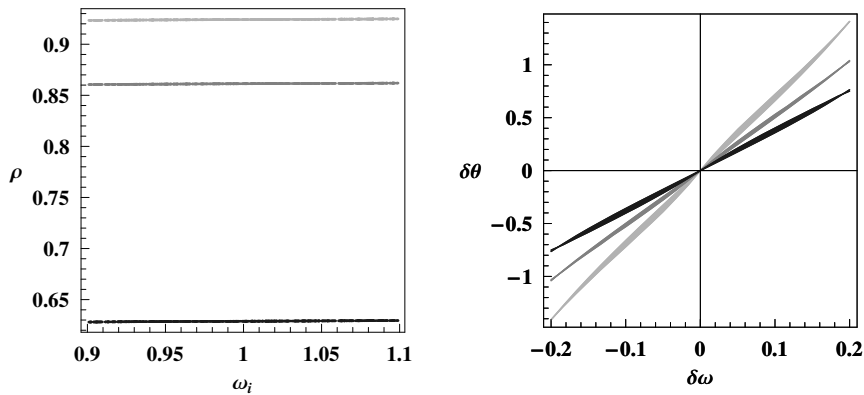,height=6cm}
\end{center}
\caption{{\small On the left: the rotation number $\rho$ plotted
versus $\omega$ for $\gamma=0.10$ and $\kappa=0.35,\,0.45,\,0.65$
(from top to bottom). On the right: $\delta\theta$ versus
$\delta\omega$ for the same values of $\gamma$ and $\kappa$
(larger slopes correspond to smaller values of $\kappa$).
\label{fig1}}}
\end{figure}

We now wish to study  the relation between the  phase shift $\delta\theta$ of a pair of
oscillators and the difference of their natural frequencies $\delta\omega$. We have
performed numerical simulations with $N=500$ oscillators with different values of
$\kappa$ and $\gamma=0.10$ corresponding to the synchronization phase. We have considered
times as large as $t=1,000$. As expected there is no dependence on the initial
conditions. On the contrary $\delta\theta$ is linearly related to
 $\delta\omega$
as shown in Fig. \ref{fig1}, where we plot $\rho_i$ versus
$\omega_i$ for various values of $\kappa$ (on the left) and
$\delta\theta$ versus $\delta\omega$ (on the right).
This dependence can be understood as follows. As $N\to\infty$, the
sum over all oscillators in~(\ref{eq1}) can be replaced by an
integral, yielding
\begin{equation}
v(\theta,t,\omega)=\omega-\sigma(t)\sin\theta
\label{eq3}
\end{equation} where
\begin{equation}
\sigma(t)=
\kappa\int_0^{2\pi}\int_{1-\gamma}^{1+\gamma}\left(1+\cos\theta\right)
p\left(\theta,t,\omega\right) g(\omega) d\omega d\theta\ .
\label{eq4}
\end{equation}Here $p\left(\theta,t,\omega\right)$ denotes the density of
oscillators with phase $\theta$ at time $t$. We consider the large
$t$ behavior to allow for synchronization; moreover we take a
temporal averaging over the common period $T$ to get rid of local
fluctuations. We get
\begin{equation}
v=\omega-\frac{1}T\int_{t}^{t+T}dt\,\sigma(t)\sin\theta(t)
\label{eq5}
\end{equation}and consider variations in $\omega$:
\begin{equation}
0=\delta\omega-\frac{1}T\int_{t}^{t+T}dt\,\sigma(t)\,
\delta\theta(t)\cos\theta(t)\ . \label{eq6}
\end{equation}Since the oscillators are synchronized
$\delta\theta(t)$ is time-independent for $t$ large enough.
Therefore
\begin{eqnarray}
\delta\omega&=&\frac{\kappa\delta\theta}T\int_{t}^{t+T}dt\int
d\omega\, g(\omega)\int_0^{2\pi}d\hat\theta\,(1+\cos\hat\theta)\,
p(\hat\theta,t,\omega)\,\cos\theta(t)=\cr
&=&\frac{\kappa\delta\theta}{T}\int_{t}^{t+T}dt\cos\theta(t) +
\frac{\kappa\delta\theta}{2T}\int_{t}^{t+T}dt\int d\omega\,
g(\omega)\int_0^{2\pi}d\hat\theta\,p(\hat\theta,t,\omega)\,
\cos[\hat\theta+\theta(t)]+\cr&+&
\frac{\kappa\delta\theta}{2T}\int_{t}^{t+T}dt\int d\omega
g(\omega)\int_0^{2\pi}d\hat\theta\,p(\hat\theta,t,\omega)\,
\cos[\hat\theta- \theta(t)] \label{eq7}\ .
\end{eqnarray}
The first two terms on the r.h.s. of (\ref{eq7}) vanish since the
integrand functions have zero temporal average.
 We get therefore
\begin{equation}\delta\omega
=\frac{\kappa\lambda}{2}\,\delta\theta, \label{eq8}
\end{equation} which is the desired linear relation between
$\delta\omega$ and $\delta\theta$. The factor $\lambda$ is
evaluated as follows:
\begin{eqnarray}
\lambda&=&\frac{1}{T}\int_{t}^{t+T}dt\int_{1-\gamma}^{1+\gamma}
d\omega g(\omega)\int_0^{2\pi}d\hat\theta\,p
\left(\hat\theta,t,\omega\right) \cos[\hat\theta- \theta(t)]= \cr
&=&\frac 1 N\sum_{j=1}^N\frac{1}{T}\int_{t}^{t+T}dt
\cos[\theta_j(t)- \theta(t)] =\frac 1
N\sum_{j=1}^N\cos[\delta\theta_j]=\cr&=&
\frac{4\gamma}{\kappa\lambda}\int^{+1}_{-1} dy\,\tilde
g\left(\frac{4\gamma\,y}{\kappa\lambda}\right)\,\cos
\left(\frac{4\gamma\,y}{\kappa\lambda}\right)\ .
 \label{eq9}
\end{eqnarray}Here $\tilde g(\delta\theta)$ is
probability distribution of $\delta\theta$. It is related to the
probability distribution of $\delta\omega $ by (\ref{eq8}). Both
the $\delta\omega $ and the $\delta\theta$ distribution functions
are derived from the $\omega $ density $g(\omega)$. If this
density is flat, as assumed here, then $\tilde g=g\star g$, i.e.
\begin{equation}
 \tilde
g(\delta \theta)=\frac{
  \kappa\lambda}{4\gamma}-\left(\frac{\kappa\lambda}{4\gamma}
  \right)^2 |\delta \theta| \ .
\end{equation}
In conclusion $\lambda$ is given by solving the equation
\begin{equation}\frac{4\gamma^2}{\kappa^2}=\lambda
\sin^2\left(\frac{2\gamma}{\kappa\lambda}\right)\,.\label{dth}
\end{equation} We
notice that in (\ref{eq8}) there is no dependence on $\omega_0$;
this dependence is in the first two terms of (\ref{eq7}) since
they  vanish only in the large $N$, large $t$ limit.

\begin{figure}[ht!]
\begin{center}
\epsfig{file=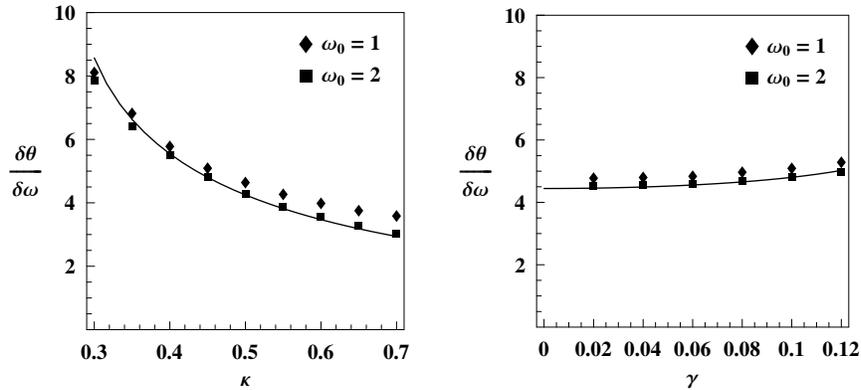,height=6cm}
\end{center}
\caption{{\small The slope
$\displaystyle\frac{\delta\theta}{\delta\omega}$ in the Winfree
model. On the left:
$\displaystyle\frac{\delta\theta}{\delta\omega}$ as a function of
$\kappa$ for two values of $\omega_0$ and $\gamma=0.1$. On the
right: $\displaystyle\frac{\delta\theta}{\delta\omega}$ as a
function of $\gamma$ for two values of $\omega_0$ and
$\kappa=0.45$. The curves are independent of $\omega_0$.
\label{fig_2}}}
\end{figure}

In Fig. \ref{fig_2} we report the slope $\delta\theta/\delta\omega$  as computed by
(\ref{dth}), as a function of $\kappa$ (with $\gamma=0.1$) on the left and  as a function
of $\gamma$ (with $\kappa=0.45$) on the right. This curve is independent of $\omega_0$.
We also report results of the numerical analysis for two values of $\omega_0$  These data
show a small dependence on $\omega_0$  \cite{nota}.

\subsection{Kuramoto model\label{kuramotomodel}} 
\begin{figure}[ht!]
\begin{center}
\epsfig{file=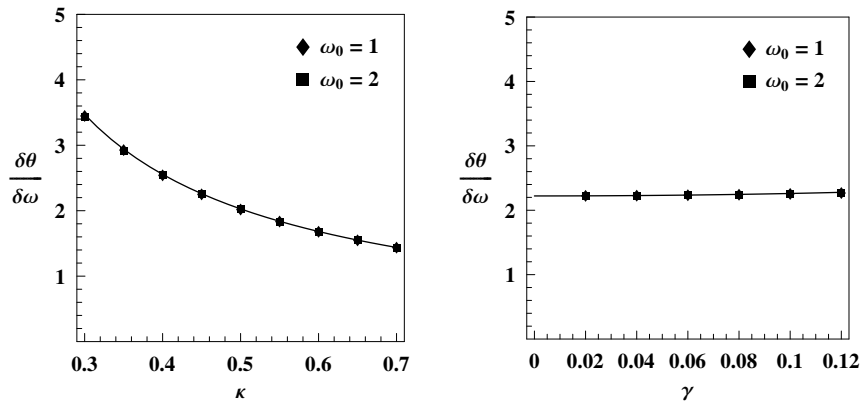,height=6cm}
\end{center}
\caption{{\small The slope
$\displaystyle\frac{\delta\theta}{\delta\omega}$ in the Kuramoto
model. On the left:
$\displaystyle\frac{\delta\theta}{\delta\omega}$ as a function of
$\kappa$ for two values of $\omega_0$ and $\gamma=0.1$. On the
right: $\displaystyle\frac{\delta\theta}{\delta\omega}$ as a
function of $\gamma$ for two values of $\omega_0$ and
$\kappa=0.45$.\label{fig_2a}}}
\end{figure}

The
analysis of the Kuramoto model produces comparable  results. The
Kuramoto model is based on the set of equations
 ($i=1,...N$)
\begin{equation}
 \dot\theta_i(t)=\omega_i+\frac{\kappa}N\sum_{j=1}^N
\sin(\theta_i-\theta_j)\ .\label{eq12}
\end{equation} The numerical results one gets  are
similar to those of fig. \ref{fig1}, with a linear dependence of
$\delta\theta$ on
 $\delta\omega$ and the rotational frequency
 $\rho_i=\omega_0$ (we use the same distribution function $g(\omega)$
 as before). The latter results follows from the fact that the phases
  $\theta_i$ are
 dynamically pulled toward the the mean phase \cite{stroreview}.
 These results  can be compared with an analytical
 treatment
  by observing that, in this case, instead of (\ref{eq7}) one gets
\begin{equation}
\delta\omega= \frac{\kappa\delta\theta}{T}\int_{t}^{t+T}dt\int
d\omega g(\omega)\int_0^{2\pi}d\hat\theta\,p(\hat\theta,t,\omega)
\cos[\hat\theta- \theta(t)] \label{eq7kuramoto} \ .
\end{equation}Due to the absence of terms analogous to the first and second terms
on the r.h.s. of (\ref{eq7}), we expect a better agreement between
numerical simulations and analytical evaluation. From
(\ref{eq7kuramoto}) we get, instead of (\ref{eq8}):
\begin{equation}\delta\omega
=\kappa\lambda\,\delta\theta \,.
\end{equation}
with $\lambda$ given by
\begin{equation}\frac{2\gamma^2}{\kappa^2}=\lambda
\left(1-\cos\frac{2\gamma}{\kappa\lambda}\right)\,.
\end{equation}
These results are reported in fig. \ref{fig_2a} together with the
results of the numerical simulations.
\section{Systolic/diastolic arterial pressure relation\label{phy}}
\subsection{Phase shifts from arterial pressure data} 
\begin{figure}[ht!]
\begin{center}
\epsfig{file=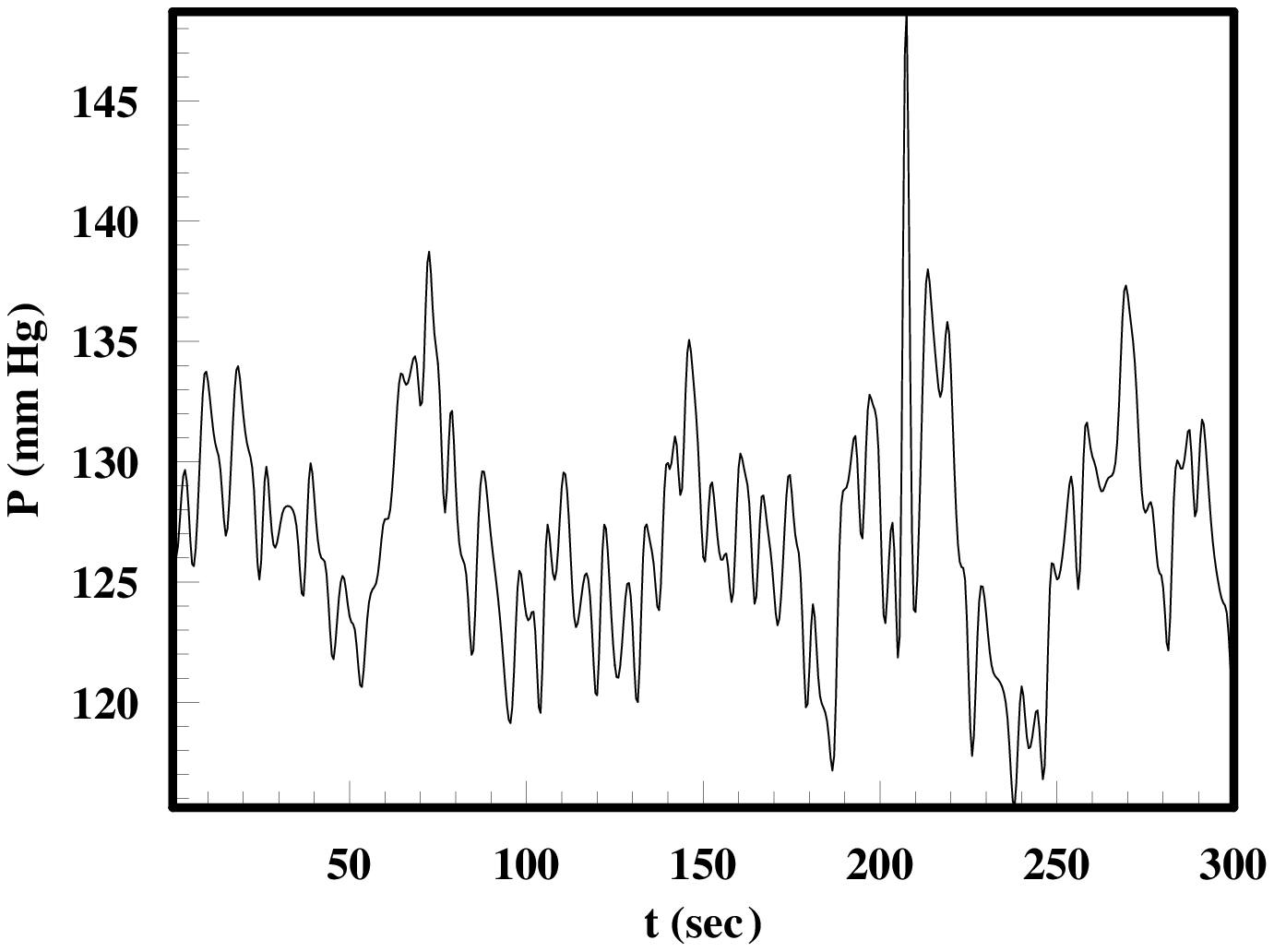,height=6cm}\epsfig{file=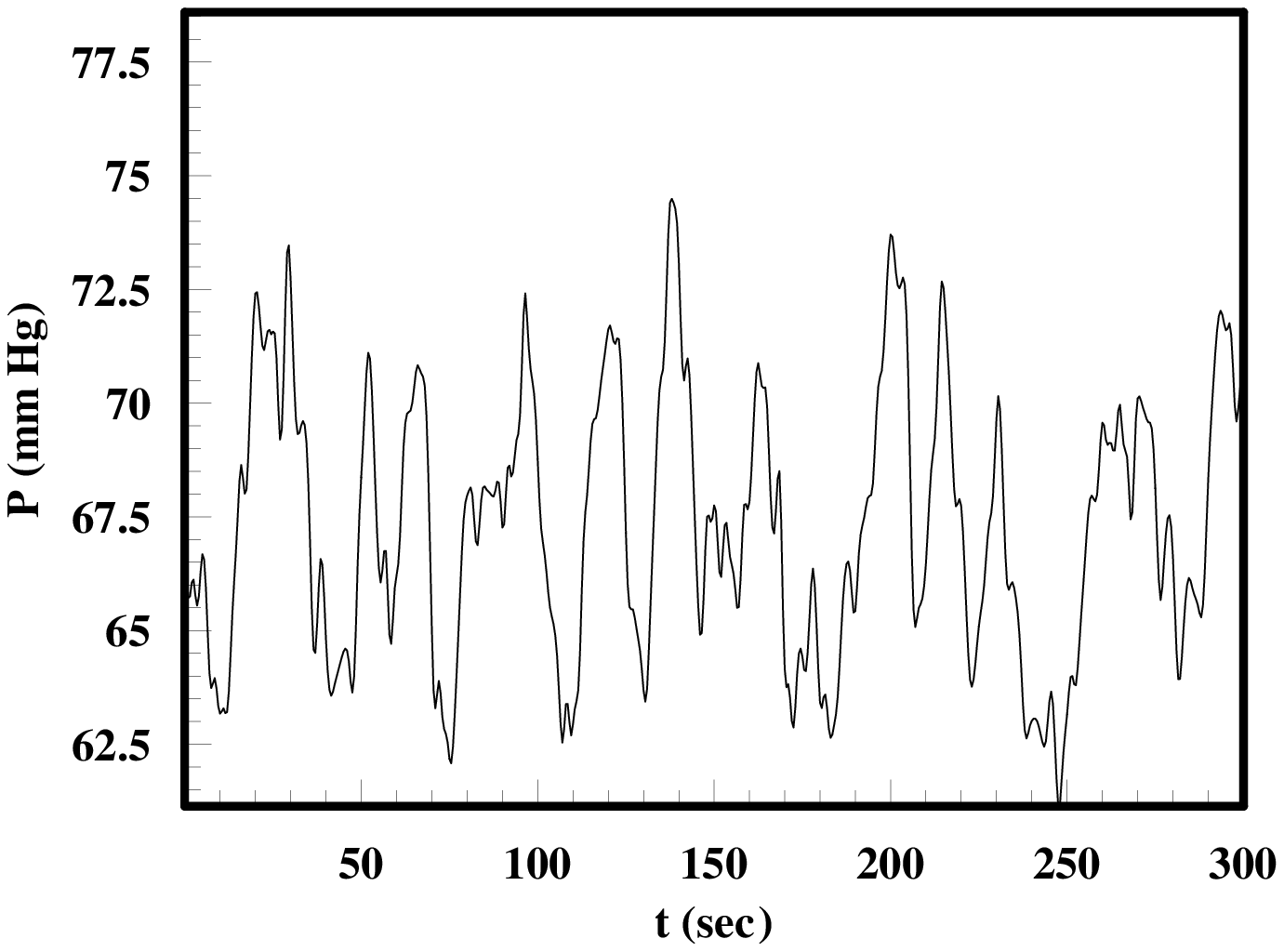,height=6cm}
\end{center}
\caption{{\small The time series $x_S(t)$ (on the left) and
$x_D(t)$ (on the right) of the systolic and diastolic arterial
pressures for one of the subjects examined. \label{fig6}}}
\end{figure}

Let us consider two time series: $x_S(t)$ and $x_D(t)$, representing systolic and diastolic
arterial pressure. Data are from a population of 47 normal
subjects (mean age+/-SD: 54+/-8 years) who underwent a 10 minutes
supine resting recording of ECG and noninvasive arterial blood
pressure (by the Finapres device), in the laboratory for the
assessment of Autonomic Nervous Sytem, S. Maugeri Foundation,
Scientific Institute of Montescano, Italy.  For each cardiac
cycle, corresponding values of SAP and DAP were computed and
resampled at a frequency of 2 Hz using a cubic spline
interpolation. In Fig. \ref{fig6} we report the time series of the
systolic arterial pressure $x_S(t)$ for one of the subjects
examined in this study.

\begin{figure}[ht!]
\begin{center}
\epsfig{file=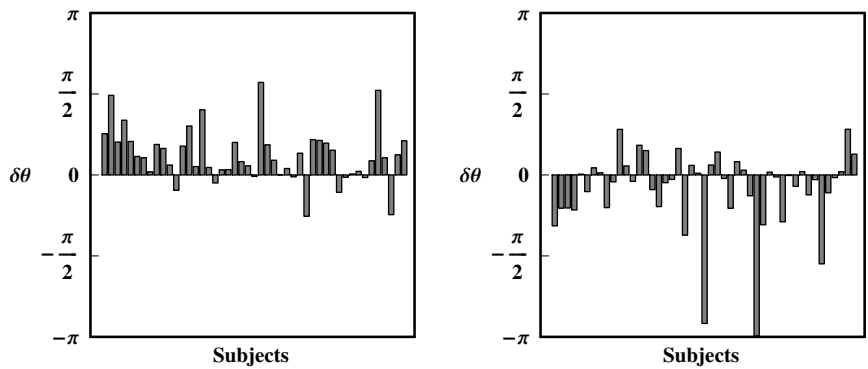,height=6cm}
\end{center}
\caption{{\small The phase shifts $ \delta\theta$ for all the 47
subjects filtered in the VLF (on the left) and HF (on the right)
bands.\label{fig7}}}
\end{figure}

 These data are analyzed by filtering in appropriate
frequency bands. We consider here three bands: Very Low Frequency (VLF) band: $
(0.01,\,0.04) $ Hz;  Low Frequency (LF) band: $ (0.04,\,0.15) $ Hz; High Frequency (HF)
band: $ (0.15,\,0.45)  $ Hz. In a previous paper \cite{pinna}, using Fourier transform
methods, occurrence of delays between SAP and DAP was investigated, and it was found that
DAP anticipates SAP in VLF (delay equal to 2.5 sec )and in LF (0.6 secs); no significant
delay was found in HF. Here we enlarge the statistical population with respect to
\cite{pinna}, and evaluate
 the phases of signals by the analytic signal technique, which allows a better estimate.
As well known SAP and DAP are highly synchronized and our data confirm this expectation.
We have used the Hilbert transform method that allows to detect phase synchronization in
noisy scalar signals \cite{tass98}. To extract a phase from the signal one considers the
Hilbert transform  of the original time series
\begin{equation}
y(t)=\frac 1 \pi
P.V.\int_{-\infty}^{+\infty}\frac{x(\tau)}{t-\tau}\,d\tau\
,\label{ht}\end{equation} where $P.V.$ denotes Cauchy principal
value. Then one forms the analytic signal
 $z(t)= x(t)+iy(t)=A(t)e^{i\phi(t)}$,
 where $A(t)=\sqrt{x^2(t)+y^2(t)}$ and $\phi(t)$ is the desired phase.
 To control the possible synchronization of two signals $x_1(t)$,
 $x_2(t)$ the following procedure is applied: the
 phases $\phi_1(t)$ and $\phi_2(t)$ are computed and
the so called {\it generalized phase
 differences} $ \Phi_{n,m}(t)=\left[m \phi_1(t)\,-\,n\phi_2(t)\right]_{mod
2\pi}$, with $n,m$ integers, are evaluated. In the present study
only $1:1$ synchronization has been examined and the two phases
$\phi_1(t)$, $\phi_2(t)$ coincide with the phases of the time
series $x_D(t),\,x_S(t)$. Phase synchronization is characterized
by the appearance of peaks in the distribution of the phase
difference. To evaluate the phase shift we have considered time
intervals characterized by a constant phase difference between the
two series:
\begin{equation}
 \delta\theta=\theta_D(t)- \theta_S(t)\ .\label{eqds}
\end{equation} We find $ \delta\theta>0$ in the VLF band, i.e. in
this  band diastolic
 pressure anticipate systolic pressure.
 Our analysis confirm the results of \cite{pinna} with a different method.
 On the other hand in the HF band (in basal conditions) the phase
 shift is negative $ \delta\theta<0$, which means that in this
 band the  systolic pressure signal anticipates the diastolic one.
 These data are reported for all the 47 subjects in fig. \ref{fig7} that shows
 on
 the left the VLF  band and on the right the HF band (we have not reported
  data in the intermediate region LF band, as
they are compatible with $ \delta\theta=0$). We estimated
$1.76\times 10^{-6}$ to be the probability that the phase shifts
in the VLF band are sampled from a distribution whose mean is less
than or equal to zero; analogously $3.0\times 10^{-2}$ is the
probability that the phase shifts in the HF band are sampled from
a distribution whose mean is greater than or equal to zero.

On a physiological basis these results mean that the different
sets of oscillators producing the time series have different
spectral properties. Leaving aside the task of a
physiologically-based modelization  we now show that the results
obtained in Section \ref{win}  can shed light on this phenomenon.
For the present application we use the Winfree model.

\subsection{Interpretation of phase shifts between related oscillatory
signals} 
\begin{figure}[ht!]
\begin{center}
\epsfig{file=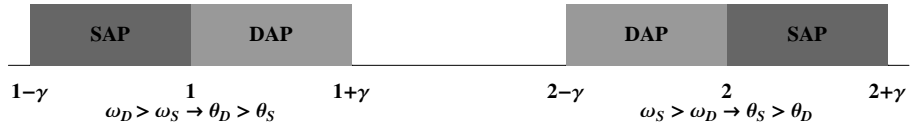,height=2cm}
\end{center}
\caption{{\small The two intervals of natural frequencies $A$ and
$B$. $A$ is  on the left and is  centered around the value $a=1$;
$B$, on the right, is centered around $b=2$. We assume that
oscillators with frequencies in the band SAP (resp. DAP)  produce
collectively the signal $x_S(t)$ (resp. $x_D(t)$), see text.
\label{bande3}}}
\end{figure}

We present here a schematic view of phase shifts
$\delta\theta$ between the  time series $x_S(t)$ and $x_D(t)$.
This picture is only qualitative and aims to reproduce the
dependence of the sign of $\delta\theta$ on the filter in
frequency power spectrum. As such, the picture is not realistic
and does not offer a physiologically-based model of the time
series; nevertheless it can shed light on oscillator dynamics
underlying them. Let us assume that the two oscillatory signals
$x_S(t)$ and $x_D(t)$ are the result of the collective behavior of
two sets of oscillators, sets $SAP$ and $DAP$ respectively. We
assume that this collective behavior produces a
 Systolic Arterial
Pressure ($SAP$) and  Diastolic Arterial Pressure ($DAP$) time
series. We assume that the oscillators in the set $SAP$ have
natural frequencies in the domain
$\omega\in(a-\gamma,a)\bigcup(b,b+\gamma)$, while frequencies for
the set $DAP$ are in the domain
$\omega\in(a,a+\gamma)\bigcup(b-\gamma,b)$. We also assume
$\gamma\le 1$. We will use as numerical values  $a=1,\,b=2$ and
$\gamma=0.1$, see Fig.\ref{bande3}. \vskip1cm

 On the
other hand the two bands $A$: $1-\gamma<\omega<1+\gamma$,  and
$B$: $2-\gamma<\omega<2+\gamma$ would model the VLF and HF
frequency bands.

\begin{figure}[ht!]
\begin{center}
\epsfig{file=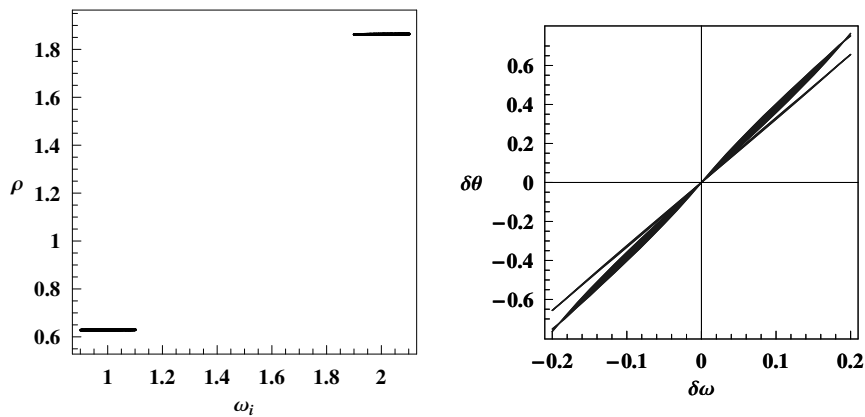,height=6cm}
\end{center}
\caption{{\small On the left: The oscillators of set $A$, with frequencies centered
around $a=1$ become synchronized with a frequency around $\omega_{VLF}=0.62$ Hz; those of
set $B$ (frequencies  around $b=2$) have a synchronization frequency $\omega_{HF}=1.88$
Hz.  On the right: The phase shift $\delta\theta$ between any pair of oscillators as a
function of the difference $\delta\omega$ between the natural frequencies of the
oscillators in the pairs. The partially overlapping lines refer to the two sets of
oscillators $A$ and $B$, which shows a weak dependence of the slope on the natural
frequencies. Numerical results refer to $N=1000$ oscillators, with $\kappa=0.65$.
\label{bands}}}
\end{figure}

Let us  assume that the coupling among the oscillators having
natural frequencies in the intervals $A$ and $B$ is modelled by
the Winfree model, i.e. by eq. (\ref{eq1}). However we assume for
the coupling \begin{equation} \kappa\to
\kappa_{ij}\,=\,\kappa\,H\left[2\gamma-|\omega_i-\omega_j|\right]\
,
\end{equation} where $H$ is the Heaviside function. By this choice there is no interaction between oscillators
in the two bands, though a weak coupling would not alter the
qualitative picture. We consider one value of $\kappa$
($\kappa=0.65$ in this case). The two sets of oscillators, one
centered around the natural frequency $\omega_0=1$ (Set $A$) and
the other around $\omega_0=2$ (Set $B$) become synchronized around
two synchronization frequencies, $\omega_{VLF}=0.62$ Hz  e
$\omega_{VLF}=1.88$ Hz see Fig.\ref{bands} (left side).
Correspondingly, on the right, we have two lines showing a linear
dependence between $\delta\omega$ and $\delta\theta$. The two
lines are almost completely superimposed, which shows a weak
dependence on the average natural frequencies of the two sets. The
interesting result however is related to our definition of VLF and
HF bands. The VLF band is the result of the collective behavior of
oscillators in the set $A$. For them $\omega_D>\omega_S$ and
therefore, on the basis of the results of Section \ref{win},
$\delta\theta=\theta_D(t)- \theta_S(t)>0$. On the other hand in
the HF band, $\omega_S>\omega_D$ and therefore $\delta\theta<0$.
This simple mechanism implies the effect of a change of sign
between the two phases when one goes from the very low frequency
to the high frequency.

One might wonder if our data also show a casual dependence between the two time series.
To address this issue we have considered the index $S(X|Y)$ that measures the nonlinear
interdependency between two time series $X$ and $Y$, as described in \cite{lehnertz}.
More precisely, from the time series ${\bf x}$ and ${\bf y}$, one reconstructs delay
vectors ${\bf x}_n=(x_n,...,x_{n-(m-1)\tau})$ and ${\bf y}_n=(y_n,...,y_{n-(m-1)\tau})$,
where $n=1,...,N$ is the time index, $m$ is the embedding dimension, and $\tau$ denotes
the time lag. Let $r_{n,j}$ and $s_{n,j}$, $j=1,...,k$ denote the time indices of the $k$
nearest neighbors of ${\bf x}_n$ and ${\bf y}_n$, respectively. For each ${\bf x}_n$, the
mean squared Euclidean distance to its $k$ neighbors is defined as
\begin{equation}
R_n^{(k)}({\bf X})={1\over k} \sum_{j=1}^k({\bf x}_n - {\bf
x}_{r_{n,j}})^2,
\end{equation}
while the {\bf Y}-conditioned mean squared Euclidean distance is defined by replacing the
nearest neighbors by the equal time partners of the closest neighbors of ${\bf y}_n$,
\begin{equation}
R_n^{(k)}({\bf X}|{\bf Y})={1\over k} \sum_{j=1}^k({\bf x}_n -
{\bf x}_{s_{n,j}})^2.
\end{equation}
The interdependence measure is then defined as
\begin{equation}
S({\bf X}|{\bf Y})={1\over N} \sum_{n=1}^N {R_n^{(k)}({\bf
X})\over R_n^{(k)}({\bf X}|{\bf Y})}.
\end{equation}
 $S(X|Y)$ is an asymmetric quantity and the
degree of asymmetry is connected to causal relationship between the two time series, in
other words if $S(X|Y)$ is much greater than $S(Y|X)$ then we may conclude that Y is
driving X. On the other hand each of these values measures by its size the degree of
interdependency, $S=0$ (resp. $S=1$) meaning minimal (resp. maximal) interdependency . We
evaluated these indexes both on the SAP and DAP time series $x_S(t)$, $x_D(t)$ and on
their phases $\theta_S(t)$, $\theta_D(t)$. In both cases the asymmetry was not
significant, which means that there is no causal relationship between SAP and DAP time
series; however the results obtained with the phases are always much greater than those
obtained with the full signals. Quantitatively, the average values for the HF band are:
$S(x_D|x_S)= 4.8\times 10^{-3}$, $S(x_S|x_D)= 4.6\times 10^{-3}$, which shows a very
small asymmetry and, at the same time, a very small interdependency. As to the phases, we
get $S(\theta_D|\theta_S)= 0.899 $, $S(\theta_S|\theta_D)= 0.901$, which on the contrary
shows a larger interdependency. Similar results are obtained in VLF and LF  bands.
Besides showing the absence of a causal relation, these results confirm that in these
systems of oscillators the main source of information on the underlying structures
resides in the dynamics of the phases.

\section{Conclusions\label{conclu}}Our results represent an
original analysis of the relation between systolic/diastolic blood
pressure, which completes  previous  studies \cite{pinna}. The
measured delays between the oscillatory components of systolic and
diastolic blood pressure time series, show a change of sign going
from low to high frequencies.We have addressed it within the
paradigm of coupled nonlinear synchronous oscillators. We have
shown, using Winfree and Kuramoto models, that once
synchronization is achieved, the phase delay between oscillators
is determined by the underlying structure and we have found a
linear relationship between oscillator  phase shifts and the
difference of their natural frequencies. We then used these
results to describe our findings, that confirm that changes in the
modulating factors of arterial pressure affect differently the
systolic and diastolic pressure values \cite{med2}.


\begin{thebibliography}{99}
\bibitem{winfree80} A.~T. Winfree, {\em The Geometry of Biological
Time\/} (Springer, New York, 1980).
\bibitem{med1} A. Malliani et
al., Circulation {\bf 84}, 482 (1991); J.K. Triedman and P. Saul,
Circulation {\bf 89}, 169 (1994); J.P. Saul, Am. J. Physiol. {\bf
261}, H153 (1991).\bibitem{med2}N. Stergiopulos et al., Am. J.
Physiol. {\bf 270}, 2050 (1996).

\bibitem{pinna}G.D. Pinna, R. Maestri, M.T. La Rovere, and A. Mortara,
IEEE Computers in Cardiology {\bf 24}, 207 (1997).
\bibitem{winfree67} A.~T. Winfree, J.~Theor. Biol. {\bf 16}, 15 (1967).

\bibitem{kuramoto}Y. Kuramoto, in
 {\it International Symposium on Mathematical
 Problems in Theoretical Physics}, Vol. 39 of {\it
 Lecture Notes in Physics},
  edited by H. Araki (Springer-verlag,
 Berlin, 1975); {\it Chemical Oscillations,
 Waves and Turbulence} (Springer-verlag,
 Berlin, 1984).\bibitem{stroreview} S.~H. Strogatz, Physica~D {\bf 143},
 1 (2000).
\bibitem{strogatz00}J.T. Ariaratnam and  S.H. Strogatz, Phys. Rev. Lett. {\bf 86}, 4278 (2001).
\bibitem{walker} T.~J. Walker, Science {\bf 166}, 891 (1969);
E.~Sismondo, {\em ibid.}, {\bf 249}, 55 (1990).
\bibitem{buck} J.~Buck, Quart. Rev. Biol. {\bf 63}, 265 (1988).
\bibitem{peskin} C.~S. Peskin, {\em Mathematical Aspects of Heart
Physiology\/} (Courant Inst. Math. Sci., New York, 1975); D.~C.
Michaels, E.~P. Matyas, and J.~Jalife, Circ. Res. {\bf 61}, 704
(1987).
\bibitem{nota}The origin of this dependence on
$\omega_0=2$ is in the first two terms of (\ref{eq7}) that only
vanish in the $N\to\infty\,,t\to\infty$ limit. As the relevance of
these terms is regulated by the relative width of frequencies
$\gamma /\omega_0$, a better agreement is obtained with
$\omega_0=2$.

\bibitem{tass98}P. Tass, M.G. Rosenblum, J. Weule, J.Khurts, A.
Pikovsky, J. Volkmann, A. Schnitzler and H.J.Freund, Phys. Rev.
Lett. {\bf 81}, 3291 (1998).
\bibitem{lehnertz} J. Arnhold, P. Grassberger, K. Lehnertz,
C.E. Elger, Physica D {\bf
134}, 419 (1999).
\end{thebibliography}
\end{document}